# Equilibrium of fluid membranes with tangent-plane order (TPO), elasticity of smectics with TPO, and dispiration asymmetry in smectics-C*

Jaya Kumar A.,[1] Buddhapriya Chakrabarti,[2] and Yashodhan Hatwalne[1]

[1]*Raman Research Institute, C.V. Raman Avenue, Bangalore 560 080, India*
[2]*Department of Mathematical Sciences, University of Durham, Durham DH1 3LE, UK*
(Dated: June 17, 2015)

Fluid membranes endowed with tangent-plane order (TPO) such as tilt- and hexatic order afford unique soft matter systems for investigating the interplay between elasticity, shape, topology, and thermal fluctuations. Using the spin-connection formulation of membrane energy we obtain equations of equilibrium together with free boundary conditions for ground states of such membranes. We extend the spin-connection formulation to smectic liquid crystals with TPO and show that for chiral smectics-C* this generalization leads to experimentally verifiable consequences for dispirations having topological indices (helicities) of the same magnitude but opposite signs.

A hairy ball cannot be combed flat without creating a hair-whorl, whereas it is possible to do so for a hairy torus. This familiar fact brings out the intimate connection between the topology of a surface endowed with local orientational (tangent-plane) order, and that of the orientational order on it [1–3]. Discovery of liquid crystalline smectic $L_{\beta'}$ phase of phospholipid membranes with TPO [4], and feasibility of hyper-swelling such phases [5, 6] motivated the exploration of the interplay between elasticity, topological defects, and thermal fluctuations in fluid membranes with TPO, pioneered by Nelson and Peliti [7]. The spin-connection formulation of [7] is particularly suited to the study of this interplay, and establishes that Gaussian (intrinsic) curvature of membranes acts as a source of disclinations (vortices in the TPO) for the mitigation of overall stress (from membrane- as well as TPO elasticity). Conversely, disclinations tend to buckle flat, deformable membranes. Positive and negative disclinations of equal strength prefer locally positive (sphere-like) and negative (saddle-like) Gaussian curvatures respectively, leading to asymmetry in their energies [8, 9]. This sub-field of soft condensed matter physics [3, 10, 11] is expanding and active [12]. In this paper we address the equilibrium of ground states of fluid membranes with TPO. In particular, we independently vary the membrane-shape without affecting the TPO on it (Fig.(1)) to minimize the elastic energy of [7]. We extend the spin-connection formulation to smectics with TPO (Fig.(2)), and qualitatively investigate the energetics of dispirations (topological defects in the chiral, thermotropic liquid crystal smectic-C* (SmC*), see below). Our result demonstrates that the generalization mentioned above can lead to new, experimentally testable consequences in the field of thermotropic [13] as well as lyotropic [14] smectics with TPO.

Smectic liquid crystals are one-dimensional "solids" composed of 2-dimensional fluid layers. Smectics with TPO exhibit a rich profusion of symmetries [15], and consequently a wide variety of topological defects in-cluding curvature defects [13, 15, 16]. Fig.(2) shows the schematic of smectic-C (SmC) that has vectorial TPO, with a description of SmC* in the figure caption. Owing to their periodicity smectics with TPO support dislocations as well as disclinations [13, 16, 17]. Within the sim-plest linear elasticity theory screw dislocations in smectics are half-helicoids, however, the Volterra construction [16] of screw dislocations in SmC* leads to frustration in the TPO (Fig.(3)) that can be healed by introducing a partial disclination. This combination of a partial disclination and a screw dislocation is called a wedge-screw dispiration [18–20]. Despite being rather exotic topological defects, dispirations have been observed in antiferroelectric SmC* using simple polarizing microscopy [15, 21].

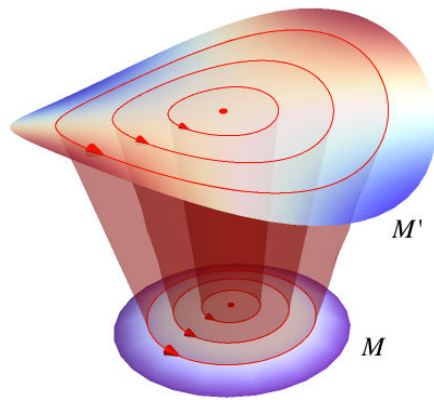

FIG. 1. (Color online) Lie dragging the TPO-field: To illustrate the idea of keeping TPO on the membrane fixed despite variation in membrane-shape we use a particularly simple example. Circular curves on the spherical, polar cap (the reference surface $M$) at the bottom are streamlines (integral curves) of the TPO-field $\hat{m}$ that rotates through $2\pi$ upon traversing a closed loop enclosing the singular polar point; they correspond to a $+1$ disclination situated at the pole. The shape of $M$ is varied (by an infinitesimal amount) to the saddle at the top ($M'$). The lightly shaded frustums of cones sandwiched between the surfaces have their apex at the center of $M$, so that the cones form loci of normals to streamlines on $M$. The streamlines on $M'$ (which also correspond to a $+1$ disclination) are obtained by dragging the streamlines on $M$ along the cones to the saddle. This ensures that there is no variation in the TPO while implementing shape variation; the streamlines, and therefore the TPO on the saddle is the same as that on the spherical cap. For Lie dragging the TPO in varying reference surfaces of arbitrary shapes, see the paragraph below (3).

Our principal results are as follows: (i) for the elastic energy (1) we obtain the first shape variation (4) together with boundary contributions (6), (ii) we show that disclination-free helicoidal membranes with a simple TPO- texture (see the paragraph following the one containing (6)) provide an exact solution to the coupled, nonlinear partial differential equations of equilibrium given by (2, 5). Moreover, we point out that for half-helicoids this solution mimics dispirations in SmC* (Fig.(3)), (iii) we adapt the spin-connection formulation for membranes to smectics with TPO (see (7,8) and the discussion that follows), and (iv) we show that the topological index that characterizes dispirations is their helicity; using (7) we find that SmC* systems prefer to have wedge-screw dispirations with positive helicity (Fig.(3)). The results (i, ii) listed above apply to membranes, interfaces and vesicles with achiral nematic-, hexatic-, and vector orders with shape-TPO coupling described by (1). For the sake of simplicity, we do not consider additional shape-TPO couplings that are specific to vector- [22], and nematic [23] order in this paper [24]. The first shape variation of (1) has been attempted earlier [25]. Our result (4) differs significantly from that of [25] (see the paragraph following the one containing (4)).

We first discuss the energetics of membranes with TPO in some detail and establish the notions and notation essential for what follows. The simplest model with TPO, the $xy$- model, has the energy density $f_{xy} \propto (\boldsymbol{\partial}\theta)^2$, where the unit $xy$- vector $\hat{\boldsymbol{m}} = (\cos\theta, \sin\theta)$, and $\boldsymbol{\partial}$ is the usual (flat space) gradient operator. To obtain the analogue of $f_{xy}$ for a deformable membrane, we set up a local, orthonormal frame $\hat{\boldsymbol{e}}_i(\underline{\sigma})$, $i = \{1, 2\}$, in the tangent plane of the membrane, where $\underline{\sigma} = \sigma^\mu$, $\mu = \{1, 2\}$, are internal coordinates on the membrane surface parametrized via the three-dimensional position vector $\boldsymbol{R}(\underline{\sigma})$. Thus $\hat{\boldsymbol{m}}(\underline{\sigma}) = (\cos\theta(\underline{\sigma}), \sin\theta(\underline{\sigma}))$ in the local Cartesian frame. In terms of the tangent vectors $\boldsymbol{t}_\mu = \partial_\mu \boldsymbol{R}$, the local Cartesian basis $\hat{\boldsymbol{e}}_i = E_i{}^\mu \boldsymbol{t}_\mu$, where components of the *vierbein* ("four legs" in German) $E_i{}^\mu$ form a $2 \times 2$ invertible matrix. We reserve Greek letters for the coordinate ($\boldsymbol{t}$-) basis, and Latin letters for the Cartesian ($\hat{\boldsymbol{e}}$-) basis. The $\hat{\boldsymbol{e}}$- basis is local; there is an $O(2)$- freedom in its choice. For deformable membranes the square-gradient

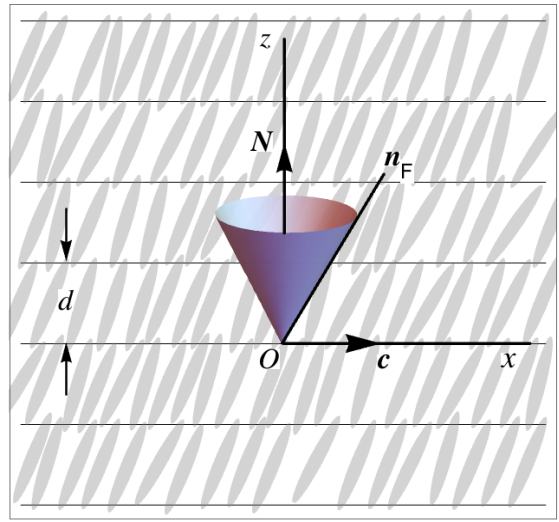

FIG. 2. (Color online) Schematics of SmC: $\boldsymbol{n}_F \equiv -\boldsymbol{n}_F$ is the unit Frank director that specifies the average orientation of molecules, $\boldsymbol{N}$ is the unit layer normal. The equilibrium layer spacing is $d$. The projection of $\boldsymbol{n}_F$ onto the layer plane, the vector $\boldsymbol{c} = (c, 0, 0)$, spontaneously breaks the continuous azimuthal symmetry. The plane spanned by $\boldsymbol{n}_F$ and $\boldsymbol{N}$ is a mirror plane with a point of inversion $O$. Structure of SmC* (not shown in the figure): SmC* has a chiral structure in which mirror- as well as inversion symmetries of SmC are lost, $\boldsymbol{c} = c(\cos(q^*z), \sin(q^*z), 0)$ in equilibrium, i.e., $\boldsymbol{n}_F$ lies on a cone with its tip on a helix with pitch $P^* = 2\pi/q^*$.

elastic energy takes the form [7]

$$F_\theta = \frac{K_A}{2} \int (\boldsymbol{\partial}\theta - \boldsymbol{A})^2 \, \mathrm{d}S, \qquad (1)$$

where the spin-connection gauge field $\boldsymbol{A}$ corrects $\boldsymbol{\partial}\theta$ so as to compensate for membrane curvature. The energy (1) is invariant under the local gauge transformation $\theta \to \theta + \eta$, $\boldsymbol{A} \to \boldsymbol{A} + \boldsymbol{\partial}\eta$. The components $A_\mu = (1/2)\,\epsilon^{ij}\,\hat{\boldsymbol{e}}_i \cdot \partial_\mu \hat{\boldsymbol{e}}_j$, where $\epsilon^{ij}$ is the totally antisymmetric unit symbol with $\epsilon^{12} = 1$. In (1), $(\boldsymbol{\partial}\theta - \boldsymbol{A})^2 = (\partial_\mu\theta - A_\mu)\,g^{\mu\nu}\,(\partial_\nu\theta - A_\nu)$, where $g^{\mu\nu}$ is the inverse of the metric tensor $g_{\mu\nu} = \boldsymbol{t}_\mu \cdot \boldsymbol{t}_\nu$. The integral is over the distorted membrane surface; the area element $\mathrm{d}S = \sqrt{g}\,\mathrm{d}\sigma^1\,\mathrm{d}\sigma^2$, where $g = \mathrm{Det}\,[g_{\mu\nu}]$. The geometry of the membrane and the topology of the $\theta$- field are connected via [7] $\boldsymbol{\nabla} \times \boldsymbol{\partial}\theta = \mathscr{S}\,\hat{\boldsymbol{n}}$, and $\boldsymbol{\nabla} \times \boldsymbol{A} = K\,\hat{\boldsymbol{n}}$, where $\boldsymbol{\nabla}$ represents the covariant gradient operator, $\hat{\boldsymbol{n}} = (\hat{\boldsymbol{t}}_1 \times \hat{\boldsymbol{t}}_2)/\sqrt{g}$ is the unit normal to the membrane, the disclination density $\mathscr{S}(\underline{\sigma}) = (2\pi/\sqrt{g}) \sum_m q_m\,\delta(\underline{\sigma} - \underline{\tilde{\sigma}}_m)$ with disclination charges $q_m$ located at $\underline{\tilde{\sigma}}_m$, and the Gaussian curvature $K(\underline{\sigma}) = \mathrm{Det}\,[K_\mu{}^\nu(\underline{\sigma})]$ is the determinant of the curvature tensor $K_{\mu\nu}$.

Next, we consider the equation of equilibrium for $F_\theta$ obtained by varying $\theta$, while keeping the membrane shape fixed [8, 9]: $(\delta F_\theta / \delta \theta) = -K_A \boldsymbol{\nabla} \cdot \boldsymbol{\mathcal{D}}\theta = 0$, where we have introduced the notation $\boldsymbol{\mathcal{D}}\theta = (\boldsymbol{\partial}\theta - \boldsymbol{A})$. The
2

Airy stress function $\chi$ defined via $\mathcal{D}^\mu \theta = \gamma^{\mu\nu} \partial_\nu \chi$, with the unit antisymmetric tensor $\gamma^{\mu\nu} = \epsilon^{\mu\nu}/\sqrt{g}$, guarantees that $(\delta F_\theta/\delta \theta) = 0$. However, $\chi$ has to obey the condition

$$\nabla^2 \chi = \mathscr{S} - K \qquad (2)$$

that ensures compatibility between the shape of the membrane and topology of the TPO embedded in it. The variational problem of minimizing $F_\theta$ also yields the free boundary condition for membranes with a boundary,

$$\hat{\boldsymbol{n}}_{(b)} \cdot \boldsymbol{\mathcal{D}} \theta = n^\mu_{(b)} \gamma_{\mu\nu} \partial^\nu \chi = 0, \qquad (3)$$

where $\hat{\boldsymbol{n}}_{(b)}$ is the unit outward normal to the boundary.

To vary the shape $\boldsymbol{R}(\underline{\sigma})$ of the membrane while keeping the $\theta$- field fixed, we set $\delta \boldsymbol{R} = \boldsymbol{t}_\mu \, \delta R^\mu_\parallel + \hat{\boldsymbol{n}} \, \delta R_\perp$, where $\delta \boldsymbol{R}_\parallel$ and $\delta R_\perp$ are respectively the variations in the tangent plane of the membrane, and along its normal. In carrying out the shape variation we need to ensure that $\delta \theta = 0$. A coordinate-independent, operational procedure for effectuating this is via Lie dragging [26] the $\hat{\boldsymbol{m}}$- field (Fig.(1)). Consider two nearby points $\boldsymbol{R}(\underline{\sigma})$ and $\boldsymbol{R}(\underline{\sigma} + \mathrm{d}\underline{\sigma})$ that are connected by $\hat{\boldsymbol{m}}$ on its integral curve in the reference membrane $M$. Upon shape variation to a configuration $M'$ along the normal to $M$, $\boldsymbol{R}'(\underline{\sigma}) = \boldsymbol{R}(\underline{\sigma}) + \hat{\boldsymbol{n}}(\underline{\sigma}) \, \delta R_\perp(\underline{\sigma})$, and $\boldsymbol{R}'(\underline{\sigma} + \mathrm{d}\underline{\sigma}) = \boldsymbol{R}(\underline{\sigma} + \mathrm{d}\underline{\sigma}) + \hat{\boldsymbol{n}}(\underline{\sigma} + \mathrm{d}\underline{\sigma}) \, \delta R_\perp(\underline{\sigma} + \mathrm{d}\underline{\sigma})$. For sufficiently small shape variations $\delta R_\perp$, the normals of points on the integral curve do not intersect (they form a congruence). The vector $\hat{\boldsymbol{m}}'$ connecting the points $\boldsymbol{R}'(\underline{\sigma})$ to $\boldsymbol{R}'(\underline{\sigma} + \mathrm{d}\underline{\sigma})$ on $M'$ is the Lie dragged version of $\hat{\boldsymbol{m}}$. Lie dragging all the integral curves on $M$ in this manner transfers the entire $\hat{\boldsymbol{m}}$- field to the varied surface.

To obtain the first variational derivative of the elastic energy (1) with respect to shape we need to find the variation of the spin connection $\delta A_\mu = (1/2) \epsilon^{ij} \delta(\hat{\boldsymbol{e}}_i \cdot \partial_\mu \hat{\boldsymbol{e}}_j)$. To this end, we first obtain $\delta \boldsymbol{e}_k$ in the $\boldsymbol{t}$- and $\hat{\boldsymbol{e}}$- bases. Using $\delta \boldsymbol{t}_\mu = \delta U_\mu{}^\nu \, \boldsymbol{t}_\nu + \delta V_\mu \, \hat{\boldsymbol{n}}$, where $\delta U_\mu{}^\nu = \nabla_\mu \delta R^\nu_\parallel + K_\mu{}^\nu$, and $\delta V_\mu = \nabla_\mu \delta R_\perp + K_{\mu\nu} \delta R^\nu_\parallel$, we have $\delta \boldsymbol{e}_k = (\delta E_k{}^\mu + E_k{}^\nu \delta U_\nu{}^\mu) \, \boldsymbol{t}_\mu + E_k{}^\mu \delta V_\mu \, \hat{\boldsymbol{n}}$ in the $\boldsymbol{t}$- basis. In the $\hat{\boldsymbol{e}}$- basis $\delta \boldsymbol{e}_k = \epsilon_k{}^l \hat{\boldsymbol{e}}_l \, \delta_\parallel + \hat{\boldsymbol{n}} \, \delta_\perp$, where $\delta_{\parallel,\perp} = \delta_{\parallel,\perp}(\underline{\sigma})$ are small variations. The term involving $\delta_\parallel$ corresponds to infinitesimal rigid anticlockwise rotation of the $\hat{\boldsymbol{e}}$- basis, and reflects the local $O(2)$ gauge freedom. We fix the gauge by setting $\delta_\parallel = 0$ (most simply done by setting $\hat{\boldsymbol{e}}_1$ parallel to $\boldsymbol{t}_1$). A comparison of the expressions for $\delta \boldsymbol{e}_k$ in the two bases then gives $\delta E_k{}^\mu = -E_k{}^\nu \delta U_\nu{}^\mu$, and $\delta \boldsymbol{e}_k = E_k{}^\mu \delta V_\mu \hat{\boldsymbol{n}}$. Substituting for $\delta \boldsymbol{e}_k$ in $\delta A_\mu$ gives $\delta A_\mu = \epsilon^{ij} E_i{}^\alpha \nabla_\alpha E_j{}^\sigma K_{\sigma \mu}$. Using $E^{\alpha i} E_i{}^\sigma = g^{\alpha \sigma}$ it is straightforward to show that $\epsilon^{ij} E_i{}^\alpha E_j{}^\sigma = \epsilon^{\alpha\sigma}/\sqrt{g} = \gamma^{\alpha\sigma}$. Thus $\delta A_\mu = \gamma^{\alpha\sigma} K_{\sigma\mu} (\nabla_\alpha \delta R_\perp + K_{\alpha\nu} \delta R^\nu_\parallel)$. This result leads to the shape variation (see [32], §1)

$$(\delta F_\theta / \delta R_\perp) = K_A \left( K^{\mu\nu} \Psi_{\mu\nu} + H \, \Phi \right), \qquad (4)$$

where the mean curvature $H = (1/2) \mathrm{Tr}\,[K_{\mu\nu}]$, and we have defined $\Psi_{\mu\nu} = \nabla_\mu \nabla_\nu \chi - (\nabla_\mu \chi)(\nabla_\nu \chi)$, $\Phi =$ $(\boldsymbol{\nabla}\chi)^2 - 2\nabla^2 \chi$. In writing (4) we have discarded the term $K_A \left( \gamma_{\mu\nu} \partial^\nu \chi \right) \left( \gamma^{\alpha\beta} \nabla_\alpha K_\beta{}^\mu \right)$ obtained through the variation, assuming nonsingular parametrization of membrane patches, so that $\gamma^{\alpha\beta} \nabla_\alpha K_\beta{}^\mu = 0$. The boundary contributions obtained from shape variation of (1) are directly included in (6).

The shape variation of $F_\theta$ has been investigated earlier [25] in the context of a model energy that ignores the mean curvature terms in the Helfrich energy for membranes and interfaces (see the next paragraph), using Green's function ("Coulomb-gas") representation of $F_\theta$ [8, 9]. The calculation of [25] is done using conformal gauge under the restriction $\mathscr{S} = 0$. The result for the variation, $\delta F_\theta = -2K_A \, H K \, \delta R_\perp$, is at variance with (4) obtained above, and does not involve the $\chi$- (and therefore, the $\theta$-) field despite the TPO-shape coupling in (1). We note that [25] implicitly uses identical parametrization for reference as well as varied surfaces; it is not clear to us how this can be implemented within the conformal gauge. We have calculated $\delta F_\theta / \delta R_\perp$ using the Coulomb-gas representation, without the restriction $\mathscr{S} = 0$ (see [32], §2) and recover our result (4) above.

In what follows we briefly describe the contributions to the energy that are common to all fluid membranes (regardless of TPO). The Helfrich energy [27] of a deformable fluid membrane is $F_H = \int [(\kappa/2)(H-H_0)^2 + \kappa_G K] \, \mathrm{d}S$, where $H_0$ is the spontaneous curvature, and $\kappa$, $\kappa_G$ are elastic constants. The contribution to the total energy from membrane surface tension $\sigma$ is $F_s = \sigma \int \mathrm{d}S$, with $\sigma \geq 0$ for stability. The case $\sigma > 0$ is particularly important for tense fluid membranes [28]. For membranes with a boundary we have to include the edge energy $F_e = \gamma \oint \mathrm{d}l$, where $\gamma$ is the coefficient of line tension, and the integral is over the boundary.

For $\mathcal{F} = F_H + F_s + F_e$ the Euler-Lagrange equations together with free boundary conditions are known [29]. For the sake of completeness and convenience, we explicitly write the full shape equation for total energy $\mathscr{F} = \mathcal{F} + F_\theta$;

$$(\delta F_\theta / \delta R_\perp) + (\delta \mathcal{F} / \delta R_\perp) = 0, \qquad (5)$$

where the first term is given by (4), and $(\delta \mathcal{F} / \delta R_\perp) = (\kappa/2)[\nabla^2 H + 2(H - H_0)(H^2 - K + H H_0)] - 2\sigma H$. To the lowest order, and for spontaneous curvature $H_0 = 0$, (5) reduces to the "nonlinear, hexatic von Kármán equation" of [8], $(\tilde{\kappa}/K_A) \nabla^4 f = (\partial_y^2 \chi)(\partial_x^2 f) + (\partial_x^2 \chi)(\partial_y^2 f) - 2(\partial_x \partial_y \chi)(\partial_x \partial_y f)$ in the Monge gauge, where $f(x,y)$ is the membrane height function, and $\tilde{\kappa} = \kappa/4$ [30]. The result of [25] does not reduce to this equation even after including the variation of the $H^2$- term from Helfrich energy. The compatibility condition (2) and the shape equation (5) form a pair of coupled, nonlinear partial differential equations of bulk equilibrium.

To obtain the contribution of $F_\theta$ to free boundary conditions resulting from shape variation we follow the methods of [29]. Describing the boundary curve in the arc-length parametrization $\boldsymbol{R}(s)$, and using the unit triad comprising the tangent to the boundary $\hat{\boldsymbol{t}}_{(b)}(s) = d\boldsymbol{R}(s)/ds$, the surface normal at the boundary $\hat{\boldsymbol{n}}(s)$, and the outward normal to the boundary $\hat{\boldsymbol{n}}_{(b)}(s) = \hat{\boldsymbol{t}}_{(b)}(s) \times \hat{\boldsymbol{n}}(s)$, we use the following notation: $\nabla_\parallel = t^\mu_{(b)} \nabla_\mu$, $\nabla_\perp = n^\mu_{(b)} \nabla_\mu$, $K_\parallel = t^\mu_{(b)} t^\nu_{(b)} K_{\mu\nu}$, $K_\perp = n^\mu_{(b)} n^\nu_{(b)} K_{\mu\nu}$, and $K_{\parallel\perp} = t^\mu_{(b)} n^\nu_{(b)} K_{\mu\nu}$. The free boundary conditions from shape variation of the total energy $\mathscr{F}$, corresponding respectively to variations along $\hat{\boldsymbol{n}}_{(b)}$, $\hat{\boldsymbol{n}}$, and $\hat{\boldsymbol{t}}_{(b)}$ are

$$K_A \left[(\boldsymbol{\nabla}\chi)^2 + (n^\mu_{(b)} \gamma_{\mu\nu} \nabla^\nu \chi)^2\right] + \hat{\boldsymbol{n}}_{(b)} \cdot (\delta\mathscr{F}/\delta\mathbf{R})\big|_{\partial M} = 0,$$
$$K_A (t^\mu_{(b)} \gamma_{\mu\nu} \nabla^\nu \chi) K_\parallel + \hat{\boldsymbol{n}} \cdot (\delta\mathscr{F}/\delta\mathbf{R})\big|_{\partial M} = 0, \text{ and}$$
$$\hat{\boldsymbol{t}}_{(b)} \cdot (\delta\mathscr{F}/\delta\mathbf{R})\big|_{\partial M} = 0, \qquad (6)$$

where the boundary contributions from the variation of $\mathcal{F}$ are: $\hat{\boldsymbol{n}}_{(b)} \cdot (\delta\mathscr{F}/\delta\mathbf{R})\big|_{\partial M} = \kappa(H-H_0)^2 + 2(\kappa_G K + \gamma k_g + \sigma)$, $\hat{\boldsymbol{n}} \cdot (\delta\mathscr{F}/\delta\mathbf{R})\big|_{\partial M} = (\kappa/2)\nabla_\perp H - \kappa_G \nabla_\parallel K_{\parallel\perp} + \gamma K_\parallel$, and $\hat{\boldsymbol{t}}_{(b)} \cdot (\delta\mathscr{F}/\delta\mathbf{R})\big|_{\partial M} = (\kappa/2) H + \kappa_G K_\parallel$. Note that in (6), $\gamma_{\mu\nu} \nabla^\nu \chi = \mathcal{D}_\mu \theta$. Thus (3) and (6) constitute the full set of free boundary conditions. For minimal surfaces ($H=0$) equations of equilibrium as well as boundary conditions simplify considerably.

We now solve the coupled equations of bulk equilibrium (2) and (5) for a disclination-free helicoid ($\mathscr{S}=0$). Using the geometrical attributes of a helicoid with pitch $p$ (see [32], §3) we find that the solution is $\chi = (1/2)\ln(\rho^2 + p^2) + [\alpha/(2\pi)]\ln(\rho + \sqrt{\rho^2 + p^2}) + \beta \implies \theta = \alpha\phi + \beta$, where $\alpha, \beta$ are arbitrary constants.

To estimate the energy of dispirations (Fig.(3)), we first adapt the spin-connection formulation for 2-dimensional membranes to set up the elastic energy of SmC* (see [20] for a "flat-space" version), and discuss the elasticity of other smectics with TPO. The energy of dispirations has been calculated in [20] using the results of [31] for the displacement field $u$ of a screw dislocation. Supplementing the Helfrich energy, the elastic energy of all smectics have a term corresponding to layer compression/dilation [13]. As for fluid membranes, TPO is inevitably coupled to the shape of smectic layering. The lowest order, isotropic, covariant coupling that is particularly important in the context of topological defects has to be of the form $F_\theta$ (1). It is straightforward to incorporate this coupling in the standard continuum elasticity of smectics with TPO. With $\gamma, H, K$, the angle $\psi$ (that replaces $\theta$ for membranes), and $\boldsymbol{A}_\perp$ functions of $(\sigma^1, \sigma^2, z)$ (where $\sigma^{1,2}$ parametrize the smectic surface, and $\hat{\boldsymbol{z}} \parallel \hat{\boldsymbol{N}}_0$, the smectic layer normal in its ground state), the elastic energy density of SmC* is

$$f_{C^*} = (B/2)\gamma^2 + (\kappa/2) H^2 + \kappa_G K + f_\psi, \qquad (7)$$

where

$$f_\psi = \frac{K_A}{2}(\boldsymbol{\partial}_\perp \psi - \boldsymbol{A}_\perp)^2 + \frac{K_N}{2}(N_\alpha \partial^\alpha \psi)^2 - h^* N_\alpha \partial^\alpha \psi. \qquad (8)$$

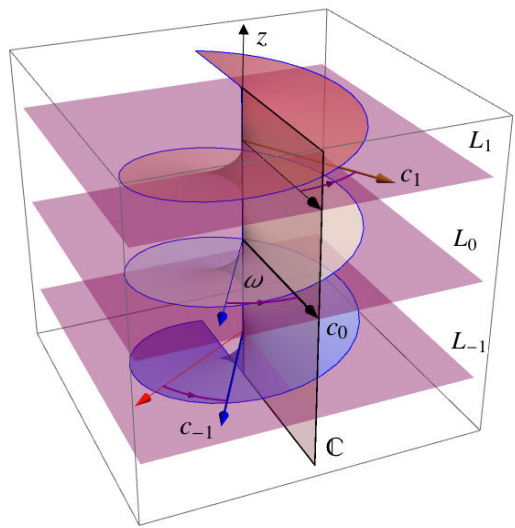

FIG. 3. (Color online) Volterra construction for a dispiration in SmC*: $L_i$ (light planar surfaces) and $c_i$ (thick arrows), $i = \{-1, 0, 1\}$, represent smectic layers and corresponding $\hat{\boldsymbol{c}}$-fields in the ground state of SmC* with inter-layer spacing $d$, and pitch $P^*$. In the right-handed laboratory frame $\hat{\boldsymbol{c}}_0 \parallel \hat{\boldsymbol{x}}$. Thin arrows on $L_i$ indicate the $\hat{\boldsymbol{c}}$- field orientation on the lower layer $L_{i-1}$. Making a vertical cut $\mathbb{C}$ (shaded rectangle) in the planar layers and successively joining the left lip of the cut on the lower layer to the right lip of the cut on the upper layer, we find that the $\hat{\boldsymbol{c}}$- fields do not match across the cut plane. To eliminate the mismatch, a wedge of angle $\omega = d/|P^*|$ *in the $\hat{\boldsymbol{c}}$- field* has to be inserted at the central singular line. This is a *partial disclination*, i.e., one that does not correspond to a symmetry operation of the ground state of SmC*. Post-relaxation, the topological construction described above leads to a wedge-screw dispiration - a screw dislocation associated with a partial disclination. To find the Burgers vector $\boldsymbol{b}$ of the screw dislocation (half-helicoid) we choose to take a circuit around the $z$- axis in a sense that ensures $\omega > 0$ (anti-clockwise in this figure). Thus $\boldsymbol{b} = d\,\hat{\boldsymbol{z}}$, and the dispiration is characterized by the helicity index $\mathbf{s}_d = (\hat{\boldsymbol{z}} \cdot \boldsymbol{b})/|P^*|$ (see [32], §4). Since $d/|P^*| \ll 1$, the overall helicity (using the right-hand rule) of the dispirated structure as defined by rotation of the $\hat{\boldsymbol{c}}$- field and the direction of Burgers vector is positive. Disclination-free textures of the form $\theta = \alpha\phi$+constant (with appropriately chosen $\alpha$) on half-helicoids are therefore structurally akin to dispirations in SmC*.

In (7), $\gamma$ describes the rotationally invariant layer compression [13, 16, 17], the next two terms describe the familiar Helfrich energy density adapted to smectics, and the term $f_\psi$ is discussed below. The energy $F_{C^*} = \int f_{C^*} \, dV$, where the volume integral is over the distorted smectic within the Eulerian formulation of elasticity [17] used here. The first term in (8) is a simple generalization of $f_\theta$ in (1), where $\boldsymbol{\partial}_\perp$ stands for the flat-space gradient operator, and $\boldsymbol{A}_\perp$ is the usual two-component spin connection, both in the tangent plane of smectic layers. The second term in (8) describes the energy cost for deviations of $\psi$ along the local unit layer normal $\hat{\boldsymbol{N}}$. The term with the pseudoscalar coefficient $h^*$ reflects the chirality of SmC* ($\psi$ is a pseudoscalar) and ensures that the equilibrium structure of SmC* is chiral, with pitch $P^* = 2\pi K_N/h^*$. To the lowest order there is no elastic coupling between $\gamma$ and gradients of $\psi$. We note that the elasticity theory described above holds for smectics-C* with $|P^*| \gg d$ (see [32], §4).

The spin-connection formulation (7) for the elasticity of SmC* is easily modified to describe elasticity of other smectics with TPO. For example, the lowest order elastic energy of SmC (isotropic in the tangent-plane, with fixed magnitude of **c**) is obtained by setting $h^* = 0$, and that of hexatic smectics by defining $\psi$ modulo $2\pi/6$, with $h^* = 0$. We have thus generalised (to the lowest order) the spin-connection formulation to smectics with TPO.

We now use the solution $\psi = \mathbf{s}_d \phi$, $\mathbf{s}_d > 0$ for the energetics of a dispiration (Fig.(3)). The full, nonlinear variational problem of minimizing (7) and solving for dispirations is formidable, and can only be tackled numerically. For the sake of analytical tractability we use standard approximations [13, 16, 17] and ignore (a) the deviation of the layer normal $\hat{\mathbf{N}}$ from $\hat{\mathbf{z}}$, and (b) the small difference between the volume elements of the reference and distorted configurations. Compensating for the components $A_\rho = 0$, $A_\phi = -1$ for the planar reference state of smectic layers in the helicoidal gauge [8], we find that the energy per unit length of a straight dispiration is

$$\frac{E}{\pi K_A} \simeq \mathbf{s}_d^2 (c_1 + \ln \lambda_2) + \mathbf{s}_d [c_2 + 2\ln(\lambda_1/\lambda_2)] + E_c + c_3, \quad (9)$$

where $\lambda_2 = (\lambda + \lambda_1)$, $\lambda_1 = \sqrt{1 + \lambda^2}$, $\lambda = L/d \gg 1$ for a system of size $L$; $c_1 = -0.88, c_2 = 1.07, c_3 \simeq 0.02$, and $E_c$ is the energy cost for destruction of smectic order near the dispiration core. The screw dislocation component does not contribute to the total elastic energy within the approximations used [13]. The "flat-space" result of [20] does not have the crucial term linear in $\mathbf{s}_d$ (9) that leads to dispiration asymmetry. From (9), $E(\mathbf{s}_d) < E(-\mathbf{s}_d)$; for $\lambda \to \infty$, $E(\mathbf{s}_d) - E(-\mathbf{s}_d) \simeq -2K_A \mathbf{s}_d$. SmC* prefers dispirations with positive helicity - a result that is amenable to experimental tests.

We thank Joseph Samuel, particularly for introducing us to Lie dragging. Discussions with V. A. Raghunathan and N. V. Madhusudana were very useful. YH acknowledges a Durham International Senior Research Fellowship, Institute of Advanced Study, University of Durham (UD), UK, where part of this work was done, and Josephine Butler College, UD, for kind hospitality.

## Supplemental material

### §1. Derivation of the shape variation of $F_\theta$:

To obtain the shape variation $(\delta F_\theta/\delta \theta)$ (equation (4) of the main text), we use $\delta g_{\mu\nu} = \nabla_\mu \delta R_{\parallel\,\nu} + \nabla_\nu \delta R_{\parallel\,\mu} - 2K_{\mu\nu}\,\delta R_\perp$ and $\delta\sqrt{g} = \sqrt{g}\,\nabla_\mu \delta R_\parallel^\mu - 2H\sqrt{g}\,\delta R_\perp$, in addition to the result for $\delta A_\mu$ obtained in the main text.

### §2. Shape variation of $F_\theta$ within the Coulomb-gas formulation:

This calculation is due to Jaya Kumar A.

The shape variation normal to the surface $(\delta F_\theta/\delta R_\perp)$ contributes to the shape equation. We do not confine ourselves to the restriction $\mathscr{S} = 0$ imposed in [SR1] (reference [25] of the main text), and use a general gauge instead of the conformal gauge. In the Coulomb-gas formulation

$$F_\theta = \frac{K_A}{2} \iint \rho(\underline{\sigma}) G(\underline{\sigma},\underline{\sigma}') \rho(\underline{\sigma}')\,\mathrm{d}S\,\mathrm{d}S', \tag{S1}$$

where $\rho(\underline{\sigma}) = \mathscr{S}(\underline{\sigma}) - K(\underline{\sigma})$, and the Green's function is defined via

$$\nabla^2 G(\underline{\sigma},\underline{\sigma}') = \delta^{(2)}(\underline{\sigma},\underline{\sigma}')/\sqrt{g(\underline{\sigma}')}, \tag{S2}$$

where $\delta^{(2)}(\underline{\sigma},\underline{\sigma}')$ is the two-dimensional Dirac delta. The area elements are $\mathrm{d}S = \sqrt{g(\underline{\sigma})}\,\mathrm{d}^2\sigma$, and $\mathrm{d}S' = \sqrt{g(\underline{\sigma}')}\,\mathrm{d}^2\sigma'$. It is convenient to use the abbreviated notation $\rho(\underline{\sigma}) = \rho$, $\rho(\underline{\sigma}') = \rho'$, $g(\underline{\sigma}) = g$, $g(\underline{\sigma}') = g'$, and $G(\underline{\sigma},\underline{\sigma}') = G$. The shape variation

$$\delta F_\theta = K_A \iint \rho' \left[G\,\delta(\rho\sqrt{g}) + (\delta G)\,\rho\,\sqrt{g}\right] \mathrm{d}^2\sigma \sqrt{g'}\,\mathrm{d}^2\sigma'$$
$$= (\delta F_\theta^{(1)} + \delta F_\theta^{(2)}), \tag{S3}$$

where the second line retains the ordering of terms in the first line.

To evaluate $\delta F_\theta$ we use the following results: $(i)$ $\delta K = H\,\nabla^2 \delta R_\perp - K_{\mu\nu}\nabla^\nu\nabla^\mu \delta R_\perp + 2HK\,\delta R_\perp$, obtained through the Gauss-Codazzi relation $K_{\mu\alpha} K^\alpha{}_\nu = 2HK_{\mu\nu} - Kg_{\mu\nu}$, $(ii)$ $\delta\mathscr{S} = 2H\mathscr{S}\delta R_\perp$, which is straightforward to calculate, and $(iii)$ the solution $\chi(\underline{\sigma}) = \int G\rho'\sqrt{g'}\,\mathrm{d}^2\sigma'$ to the compatibility condition $\nabla^2 \chi = \mathscr{S} - K = \rho$. We note that $\delta\mathscr{S}$ in $(ii)$ above is to be interpreted in a coarse-grained sense, so that $\mathscr{S}$ is the disclination density corresponding to a continuous distribution of disclinations [SR2].

Integrating $\delta F_\theta^{(1)}$ by parts and ignoring boundary contributions, we have

$$\delta F_\theta^{(1)} = K_A \int (K_{\mu\nu}\nabla^\nu\nabla^\mu \chi - 2H\nabla^2\chi)\,\delta R_\perp\,\sqrt{g}\,\mathrm{d}\sigma. \tag{S4}$$

For the evaluation of $\delta F_\theta^{(2)}$ we first vary (S2) to get

$$\delta\nabla^2 G(\underline{\sigma},\underline{\sigma}') = -\frac{\delta^{(2)}(\underline{\sigma},\underline{\sigma}')}{2g(\underline{\sigma}')}\,\delta\sqrt{g(\underline{\sigma}')}, \tag{S5}$$

where $\nabla^2 G = (1/\sqrt{g})\,\partial_\mu(\sqrt{g}\,g^{\mu\nu}\,\partial_\nu G)$. Using $\delta\sqrt{g} = -2H\sqrt{g}\,\delta R_\perp$, and $\delta g^{\mu\nu} = 2K^{\mu\nu}\sqrt{g}\,\delta R_\perp$, we obtain

$$\nabla^2 \delta G = \frac{1}{\sqrt{g}}\partial_\mu\left[\sqrt{g}\,(Hg^{\mu\nu} - K^{\mu\nu})(\partial_\nu G)\delta R_\perp\right]. \tag{S6}$$

The result (S6) above implies that

$$\nabla^\mu \delta G = (Hg^{\mu\nu} - K^{\mu\nu})(\partial_\nu G)\,\delta R_\perp, \tag{S7}$$

apart from the curl of a well-behaved vector field. Using (S7) in conjunction with the compatibility condition, its solution (see the text preceding (S4)), and integrating by parts we get

$$\delta F_\theta^{(2)} = K_A \int \left[ H(\boldsymbol{\nabla}\chi)^2 - (\nabla^\mu \chi) K_\mu^\nu (\nabla_\nu \chi) \right] \delta R_\perp \sqrt{g}\, d\sigma. \tag{S8}$$

Adding up the contributions $\delta F_\theta^{(1)}$ and $\delta F_\theta^{(2)}$, we recover the result (4) of the main text.

The shape variation in [SR1] is done entirely within conformal gauge, *while retaining identical conformal parametrizations of the reference surface as well as the varied surface.* We believe that the calculation of [SR1] (see the equations for $\delta W$ ($\delta F_\theta$ in our notation) at the bottom of the second page, and the top of the third page of this reference) therefore accounts merely for the variation of the area element $\delta(dA) = -2H\epsilon\, dA$ (note that $\epsilon = \delta R_\perp$, and $dA = dS = \sqrt{g}\, d^2\sigma$ in our notation), unsurprisingly culminating in the result $\delta W = -2KH\epsilon$ (recall that $\delta\sqrt{g} = -2H\sqrt{g}\,\delta R_\perp$ in our notation).

### §3. Geometry of a helicoid:

For a helicoid parametrized by the position vector $\boldsymbol{R} = (\rho\cos\phi, \rho\sin\phi, [p/(2\pi)]\,\phi)$, the components of the metric tensor are $g_{\rho\rho} = 1$, $g_{\rho\phi} = g_{\phi\rho} = 0$, $g_{\phi\phi} = \rho^2 + p^2$, the curvature tensor has components $K_{\rho\rho} = K_{\phi\phi} = 0$, $K_{\rho\phi} = K_{\phi\rho} = -p/\sqrt{g}$, the mean curvature $H = 0$, the Gaussian curvature $K = -p^2/g^4$, and the components of spin-connection are $(A_\rho, A_\phi) = (0, -\rho/\sqrt{g})$.

### §4. Dispirations in SmC*:

For most SmC* materials (considered in this paper), the pitch $|P^*| \simeq 10^3\, d$ [SR3]. However, there are some ferrielectric chiral smectics with $|P^*| \lesssim 10\, d$ [SR4]. The elasticity theory for SmC* discussed in the main text, and our results for dispirations do not apply to these short-pitched systems.

In what follows, we focus on the characterization of dispirations based upon their helicity.

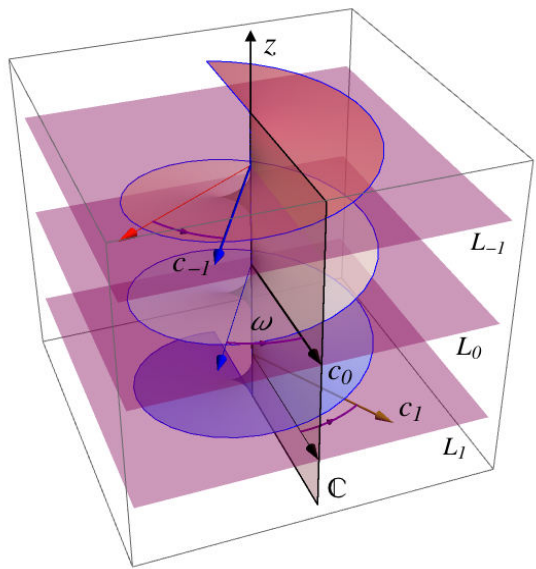

FIG. SF1. (Color online) Dispiration with negative helicity.

Because of the chirality of SmC*, dispiration lines can be assigned an unambiguous orientation and sign for $\mathbf{s}_d$ based on their helicity. This is not the case for screw dislocations in achiral smectics such as SmA or SmC. Fig. (3) of the main text shows a right-handed screw dislocation, with mismatch angle $\psi(L_i) - \psi(L_{i-1}) = \omega > 0$. To find the disclination strength we choose any curve on the half-helicoid that goes around its singular core without crossing it, and is orientated such that $\omega > 0$ in the (right-handed) cylindrical polar, laboratory frame. In doing so we are using the inherent chirality of the $\hat{c}$- field to fix the orientation of the curve. For the dispiration shown in Fig. (3) the curve is oriented anticlockwise. Along such a curve $\int_{\phi_0}^{\phi_1} \mathrm{d}u$, where $\phi_1 = \phi_0 + 2\pi$, and $\phi_0$ is an arbitrary constant, corresponds to $\boldsymbol{b} = \oint \mathrm{d}u = \hat{\boldsymbol{z}}\, d$ in the ground-state, reference SmC* lattice. Using the right-hand rule, we can then define the helicity of a dispiration using the sense of rotation of $\hat{\mathbf{c}}$ on it, and the direction of the Burgers vector. Since the mismatch angle $\omega \ll 2\pi$ ($|P^*| \gg d$), the helicity of the dispiration shown in Fig. (3) is positive.

Fig. (SF1) shows a dispiration with negative helicity. As before, choosing the orientation of any curve going around the singular $z$- axis such that $\omega > 0$ in the right-handed laboratory frame (clockwise in this figure), we find that $\boldsymbol{b} = -\hat{\boldsymbol{z}}\, d$. Thus the topological index appropriate for representing a dispiration is its helicity $\mathbf{s}_d$, as defined in the main text.

The dispiration shown in Fig. (3) is energetically favored over that of Fig. (SF1).

---